\documentclass[twoside,12pt]{article}
\usepackage{epsfig}

\newcommand{\be}{\begin{equation}}
\newcommand{\ee}{\end{equation}}
\begin{document}

\title{ \vspace{1cm} Neutrinos from CNO cycle at the present epoch of the solar
neutrino research}
\author{A.\ Kopylov,$^1$ V.\ Petukhov$^1$
\\
$^1$Institute for Nuclear Research of Russian Academy of Sciences}
\maketitle

The CNO cycle suggested by Hans Bethe \cite{1} and Carl von
Weizsackker \cite{2} contributes only a small fraction to the energy
generated in the Sun. It is the main source of energy for more
massive stars. However, the study of CNO cycle is now very important
by several reasons and future experiments will certainly aim at this
task. In a way, the unknowns related to CNO cycle constitute the
last missing chain to compose the total picture of the energy
generation in the Sun. The study of solar neutrinos achieved a great
progress during last 40 years. The discovery of neutrino
oscillations, the positive sign of $\delta $m$^{2}_{12}$ from the
observed MSW effect in the solar matter, the measurement of the
parameters of the neutrino oscillation in the solar neutrino
experiments \cite{3} - \cite{9} plus KamLAND \cite{10} -- these are
the major milestones of this study. We are approaching the era of
precision measurements in the solar neutrino research. But there is
a missing chain in this construction. Let's look at the equation of
the balance for the solar energy:

\be
0.918f_{pp}+ 0.002f_{pep}+ 0.07f_7+ 0.01f_{CNO} = 1 \label{eq:1}
\ee

Here the fluxes $f_i$ refer to the ratio of the generated fluxes to
the ones calculated by the standard solar model and the coefficients
by $f_i$ are the weights of the corresponding nuclear reactions in
the total energy generated in the Sun.

This~(\ref{eq:1}) can be rewritten as

\be 0.918f_{pp}+ 0.002f_{pep}=1 - 0.07f_7- 0.01f_{CNO} \label{eq:2}
\ee

The physical meaning of the~(\ref{eq:2}) is simple: the uncertainty
of the pp-flux (with the associated flux of pep-neutrinos directly
connected with the first one) is determined by two unknowns -- the
flux of beryllium neutrinos $f_7$ and the flux of CNO-neutrinos (so
far we do not differentiate here the flux $f_{13}$ from the decay of
$^{13}$N and the flux $f_{15}$ from the decay of $^{15}$O). The flux
$f_7$ has been measured recently in Borexino experiment \cite{11}
with the uncertainty of about 10{\%}. One can see from the
~(\ref{eq:2}) that 10{\%} uncertainty in the determination of $f_7$
can be transferred into 0.7{\%} uncertainty for pp-neutrinos. The
aim for the future of the Borexino is 5{\%} uncertainty. This will
result in 0.35{\%} uncertainty for pp-neutrinos. To get the same
contribution from CNO neutrinos the flux $f_{CNO}$ should be
measured with the uncertainty 35{\%}, i.e. 7 times higher then that
of Borexino as it follows just from the ratio of the corresponding
weights in~(\ref{eq:2}). Only then one can claim that the
uncertainty of pp-neutrinos is less then 1{\%}. This can be
summarized in the following statement: the contribution of the CNO
cycle to the solar luminosity is small but unknown, to get precision
in the evaluation of pp-flux one needs to measure $f_{CNO}$ to find
out \textit{exactly how small} this contribution is. Quite
appropriate question is -- why it is important to get a very precise
value for the flux of pp-neutrinos generated in the Sun? Future
precise measurements of the flux of pp-neutrinos can be compared
with the value found from~\ref{eq:2}. If there is no match, i.e. the
observed flux turns out to be smaller than one found
from~(\ref{eq:2}) this can be interpreted as a presence of still
unknown (hidden) source of solar energy. Other possibility -- is the
existence of sterile neutrinos by which some part of energy is
leaked out into the channel which can't be directly observed in
experiment. The only indication will be the observed deficit of
pp-neutrinos. Obviously, the sensitivity of these experiments is
directly connected with the precision obtained in measurements of
$f_7$ and $f_{CNO}$.

In evaluating the contribution of the CNO cycle to the solar energy
one should be very cautious with details. Here it is worth to
emphasize that when we mention CNO neutrinos we should take into
consideration that in CNO cycle two kinds of neutrinos are
generated: neutrinos from the decay of $^{13}$N and of $^{15}$O.
There are also neutrinos from the decay of $^{17}$F but their flux
is much lower, so far let's not take this into consideration. The
isotope $^{13}$N is generated in (p,$\gamma )$ reaction on $^{12}$C
, and $^{15}$O - on $^{14}$N. These are two branches of the main
loop of the CNO cycle (see Fig.1). Usually the conjecture is made
that CNO cycle is in a stationary state, i.e. the nuclear rates of
left and right branches of the Loop I presented on Fig.1 are nearly
equal. But in reality it may turn out to be different. It well may
be that due to some instabilities in the interior of the Sun (for
example $^3$He instability etc well discussed since long time ago)
the nuclear rate of the left branch of the Loop I is substantially
higher than the one of the right branch. The energy released in the
thermonuclear reactions of CNO cycle in this case will be
substantially different than SSM suggests. In other words: the
contribution of CNO cycle to the total energy produced in the Sun
depends not only on the heavy element abundance in the solar matter
but also on the degree of the deviation from the stationary state of
the very CNO cycle. When we discuss what is really excluded by
experimental data we should also take this into consideration. In
fact this means that it is necessary to measure both neutrino
fluxes: $f_{13}$ and $f_{15}$ and look at what energy is produced in
left and right branches of the Loop I.

\begin{figure}[h]
\begin{center}
\begin{minipage}[t]{8 cm}
\epsfig{file=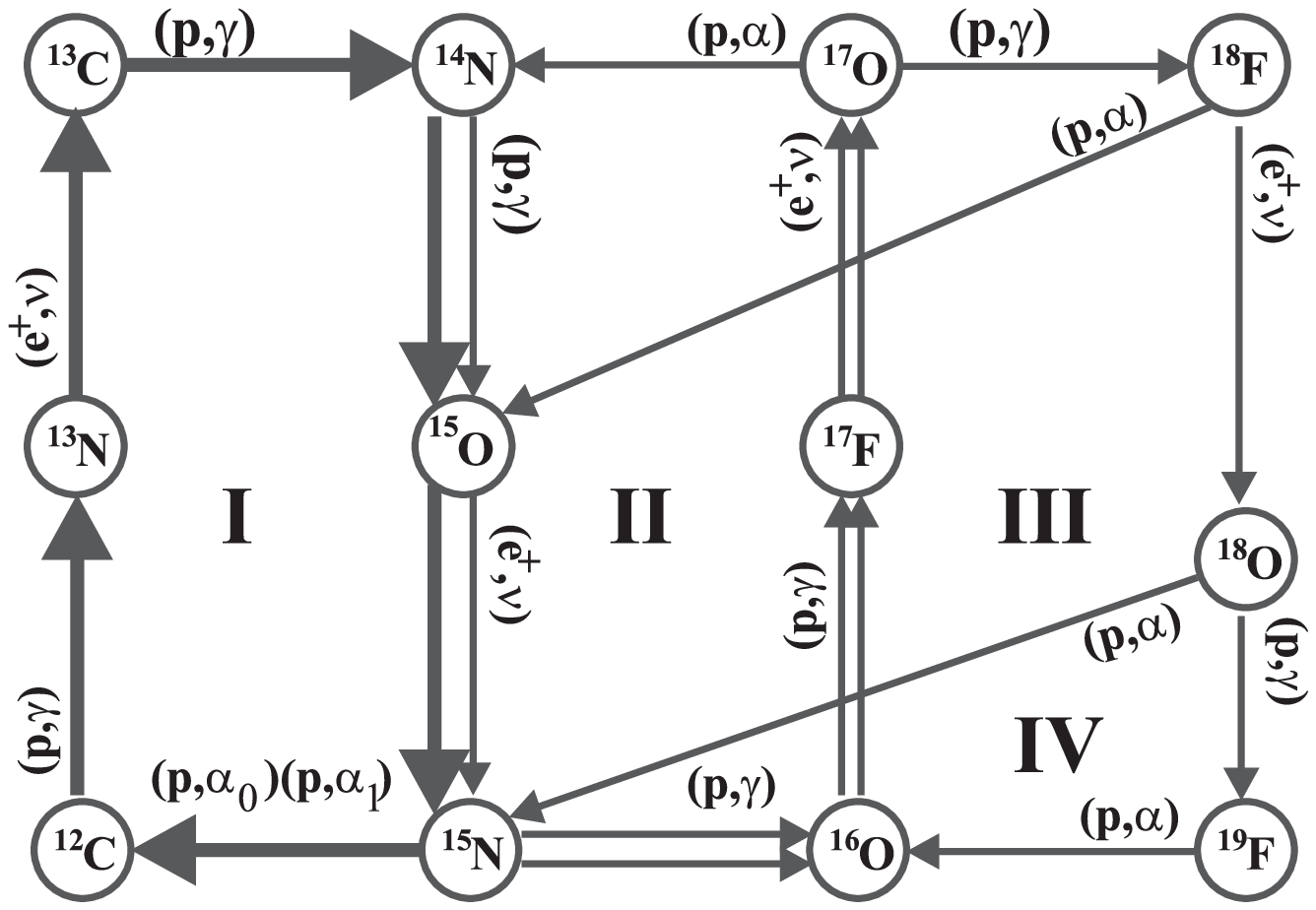,scale=0.5}
\end{minipage}
\begin{minipage}[t]{16.5 cm}
\caption{CNO-cycle.\label{fig1}}
\end{minipage}
\end{center}
\end{figure}

The isotopes $^{12}$C and $^{14}$N are used as the catalysts and are
not spent during running of the CNO cycle. But in reality the
picture is not quite symmetric. During running of CNO cycle the
isotope $^{12}$C has been burned in the thermonuclear reaction as a
fuel abundant in the material of the initial protostellar cloud. As
a result of this process at the initial stage of the formation of
the Sun the isotope $^{14}$N has been accumulated till the level
which corresponds to the stationary level when the reaction rates
for the generation of $^{12}$C and $^{14}$N nearly equal. At this
stage the flux $f_{13}$ and the flux $f_{15}$ are nearly equal.
There is some region where the temperature is high enough for the
(p,$\gamma )$ reaction on $^{12}$C but not high enough for the one
on $^{14}$N. This is why the flux $f_{13}$ is always a bit higher
than the flux $f_{15}$. But what will be in a non stationary case
when two branches of the main loop are a bit ``out of order''? Is it
possible at all that some non stationary case can be realized? It
has been shown in \cite{12}, \cite{13} that some mild mixing is
allowed when there's a small influx of fresh material from external
layers to the core of the Sun which increases substantially the
abundance of $^{12}$C in the core and, as a consequence, the flux
$f_{13}$ and, as a consequence, the energy yield from left branch of
the Loop I of the CNO cycle. The calculations show that mixing which
will not come in conflict with the results of helioseismology can
increase the flux $f_{13}$ by 50{\%} and even more and,
consequently, the energy produced by left branch also by 50{\%} and
even more. The current experimental program should be ready to this
outcome.

What possibilities do we have to measure the neutrino fluxes
generated in CNO cycle? Fig.2 shows the energy spectra of recoil
electrons from ($\nu $e$^{-}$) scattering for the case of ideal
energy resolution.

\begin{figure}[h]
\begin{center}
\begin{minipage}[t]{8 cm}
\epsfig{file=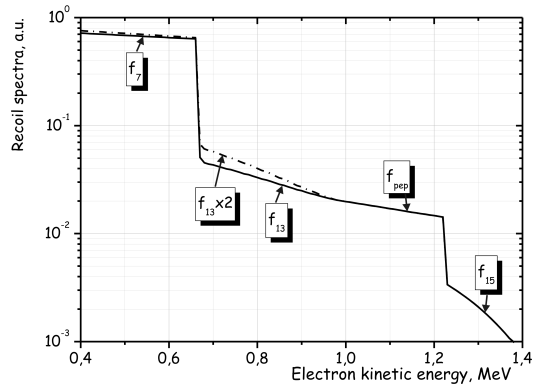,scale=0.3}
\end{minipage}
\begin{minipage}[t]{16.5 cm}
\caption{The differential energy spectra for recoil electrons for
SSM and for the model with the increased flux $f_{13}$.\label{fig2}}
\end{minipage}
\end{center}
\end{figure}

One can see from this figure that the effect from CNO neutrinos is
composed only from small fragments on top of the relatively big
areas for $^{7}$Be- and {\it pep}-neutrinos. The energy resolution
of the detectors smears the picture. The background from the
internal radioactivity is another factor, making difficult the task
of the extraction of the signal from CNO neutrinos from the measured
energy spectrum. In the meantime, the very signal from $^{7}$Be- and
pep-neutrinos is obtained by a certain conjecture about the probable
fluxes $f_{CNO}$. Then how to extract the CNO fluxes? Deconvolution
from the shape of the energy spectrum will hardly bring result in
view of all complications mentioned earlier. One possibility is to
use detectors with different sensitivities to different fluxes. For
example, if to use detectors with different thresholds of the
detection of solar neutrinos, this principally can solve the
problem. This can be both electronic and radiochemical detectors. A
good example of the first one is a LENS detector \cite{14}, of the
second one -- a lithium detector \cite{13}, \cite{15}. In future it
is quite perspective to use detectors with a large volume of
liquefied noble gases as scintillators (projects XMASS \cite{16},
CLEAN \cite{17}) to study dark matter and solar neutrinos. The
interesting possibilities are suggested also by newly developed
detector with a $^{100}$Mo target \cite{18}. These experimental
programs have good prospects for new discoveries. The Sun as a
powerful source of neutrinos of different energies will always
attract attention of experimentalists. The detectors of new
generation in newly developed underground laboratories will provide
new possibilities for high precision measurements. These two factors
present certain guarantees that study of solar neutrinos will be for
quite a long time a very fruitful field for new discoveries.

This work was supported by RFBR grant {\#}07-02-00136A and by a
grant of Leading Scientific Schools of Russia {\#}959.2008.2.

\end{document}